\documentclass[a4paper, reprint, oneside, twocolumn]{revtex4-1}

\usepackage[english]{babel}
\usepackage[utf8]{inputenc}
\usepackage[labelfont=up]{subcaption}
\usepackage[x11names]{xcolor}
\usepackage{float}
\usepackage{mathtools}
\usepackage[top=1.in, bottom=1.in, left=0.5in, right=0.5in]{geometry}
\usepackage{amsthm, amssymb}
\usepackage{graphicx}
\usepackage{lipsum}
\usepackage{bm}
\usepackage[pdftex, pdftitle={Article}, pdfauthor={Author}]{hyperref}
\usepackage{wasysym}
\usepackage{enumitem}
\usepackage{appendix}
\hypersetup{colorlinks=true, citecolor=black, linkcolor=black, urlcolor=blue}

\definecolor{brickred}{rgb}{0.8, 0.25, 0.33}

\begin{document}

    \title{Modelling Singularities in Macroevolution}
    \author{Alessandro Bellina$^{1, 2, 3}$}
    \author{Giordano De Marzo$^{1, 4, 5}$}
    \author{Vittorio Loreto$^{2, 3, 1, 5}$}

    \affiliation{$^1$Centro Ricerche Enrico Fermi, Piazza del Viminale, 1, I-00184 Rome, Italy.\\
    $^2$Dipartimento di Fisica Universit\`a ``Sapienza”, P.le A. Moro, 2, I-00185 Rome, Italy.\\
    $^3$Sony Computer Science Laboratories Rome, Joint Initiative CREF-SONY, Piazza del Viminale, 1, 00184, Rome, Italy.\\
    $^4$University of Konstanz, Universitaetstrasse 10, 78457 Konstanz, Germany\\
    $^5$Complexity Science Hub Vienna, Metternichgasse 8, 1030, Vienna, Austria.}
    
    \date{\today} 
    
\begin{abstract}
Macroevolutionary dynamics often display sudden, explosive surges, where systems remain relatively stable for extended periods before experiencing dramatic acceleration that frequently exceeds traditional exponential growth. This pattern is evident in biological evolution, cultural shifts, and technological progress and is often referred to as the emergence of singularities. 
Despite their widespread occurrence, these explosions arise from distinct underlying mechanisms in different domains. In this context, we present a unified framework that captures these dynamics through a theory of combinatorial innovation. Building on the Theory of the Adjacent Possible, we model macroevolutionary change as a process driven by recombining pre-existing elements within a system. 
By formalising these qualitative insights, we provide a mathematical structure that explains the emergence of these explosive phenomena, facilitates comparisons across different systems, and enables predictive insights into future evolutionary trajectories. Moreover, by comparing discrete and continuous formalisations of the theory, we emphasise that the occurrence and observation of these presumed singularities should be carefully considered, as they arise from the continuous limit of inherently discrete models.
\end{abstract}
\maketitle

\section{Introduction}
        
Many systems—ranging from biological~\cite{markov2007phanerozoic, grinin2015modeling, panov2004avtomodel} and social~\cite{von1960doomsday, kapitza1996phenomenological, korotayev2013globalization} to economic~\cite{lepoire2013potential, korotayev2005compact}—share a striking pattern of explosive growth and innovation. When we look at metrics like world population size~\cite{kremer1993population}, average GDP per capita~\cite{korotayev2009compact}, or historical milestones~\cite{kurzweil2004law, lepoire2015interpreting}, the pace of increase over time is undeniable. These systems often grow at rates that surpass traditional exponential models~\cite{korotayev202021st, steffen2015trajectory}, instead following hyperbolic functions~\cite{korotayev2006world, panov2005scaling}. This brings us to the concept of the 'singularity,' where growth theoretically approaches infinity in a finite time. In real-world systems, this singularity can be interpreted as a harbinger of profound transformations or disruptive events on the horizon. While hyperbolic models fit the data impressively~\cite{korotayev2009compact, panov2020singularity}, their true power and ability to explain real-world phenomena remain elusive. Additionally, the microscopic mechanisms driving such rapid growth are poorly understood, often leading to applying the same mathematical models to systems with very different behaviours.
        
The theory of "singularity" is primarily phenomenological. Many innovation-driven systems have been effectively interpreted through combinatorial processes~\cite{koppl2018simple, cortes2022biocosmology}. In this framework, innovation and growth emerge from recombining elements within the system~\cite{kauffman2000investigations, koppl2023explaining}, similar to assembling existing parts to create new tools. The Theory of the Adjacent Possible (TAP)~\cite{cortes2022tap} proposes that systems can expand into their "possible" space, which consists of elements that are not yet realised but are just one step away from being created ~\cite{kauffman2000investigations}. Models based on this concept, such as the TAP equation, have been used to describe the growth of the world population~\cite{koppl2018simple}, GDP~\cite{koppl2023explaining}, and technological innovations~\cite{sole2016singularities}. Moreover, the concept of the Adjacent Possible has also been applied within the framework of Urn Models~\cite{tria2014dynamics,di2025dynamics}, which reproduce key features of novelties dynamics, such as Zipf's, Heaps' and Taylor's laws~\cite{tria2018zipf,loreto2016dynamics}.
        
The combinatorial approach effectively replicates empirical data on a qualitative level~\cite{koppl2018simple, steel2020dynamics} but often lacks a more quantitative and mathematical analysis. In this work, we aim to reconcile the hyperbolic growth observed in natural systems and the phenomenological theory of "singularity" with the microscopically grounded theory of combinatorial innovation. This allows us to distinguish between seemingly similar systems yet governed by different underlying processes. We demonstrate the versatility of our framework by applying it to a wide range of scenarios, including world population~\cite{korotayev2006world, kapitza1996phenomenological}, biological and cosmological phase shifts~\cite{kurzweil2005singularity, korotayev2020twenty, panov2004avtomodel}, GDP per capita~\cite{korotayev2005compact}, and US patent records~\cite{youn2015invention}.
     
The modelisation we propose also sheds light on the concept of "singularity", which is often interpreted as a point where the extreme growth rate would lead to a qualitative change in the system's behaviour~\cite{kurzweil2004law, panov2020twenty, korotaev2006introduction, modis2013singularity, lepoire2020exploring}. We argue that in real-world systems, which are discrete and finite, such a concept must be carefully considered, particularly when making predictions.

\section{The TAP equation: a paradigm for combinatorial growth}

        The idea that innovations and growth arise as a combinatorial process is closely connected to the Theory of the Adjacent Possible (TAP) proposed by Kauffman et al.~\cite{kauffman2000investigations, cortes2022tap}. The concept underlying the TAP is that the range of potential future developments is generated by the existing elements within the system, creating a space of possibilities where novel elements emerge from combinations of pre-existing ones. This framework applies to both physical entities (e.g., molecules, genes, tools) and conceptual ones (e.g., patents, songs, ideas).
                
        This space of possibilities, known as the Adjacent Possible, contains all the unexplored elements that are just one step away from realisation~\cite{kauffman2000investigations}. Mathematically, this concept led to simple yet powerful models known as the TAP equations. If we denote by $M_t$ the number of elements in the system at time $t$, their evolution is governed by the number of combinations that can be formed from these elements:
        \begin{equation}
           M_{t+1} = (1-\Sigma)M_t + \sum_{i=1}^{M_t} A_i \binom{M_t}{i}
           \label{eq:TAP}
        \end{equation}
        Here, $A_i$ represents the coefficients that account for the probability of realisation of each combination, with typically decreasing values reflecting the increasing difficulty of forming longer combinations. For simplicity, we assume no extinction, setting $\Sigma = 0$. We consider all terms in the summation starting from $i=1$, although some versions of the TAP equation start from $i=2$. These models are known for their explosive behaviour, with growth rates exceeding exponential ones. Models based on the TAP concept have been used to describe the growth of the world population~\cite{koppl2018simple} and GDP~\cite{koppl2023explaining}. A similar approach has been explored by Solé et al.~\cite{sole2016singularities}, who proposed a generalised model of technological evolution. Their work incorporates ageing effects to study how recombinatorial dynamics influence innovation rates and prevent singularities in some cases. These examples emphasize the versatility of combinatorial frameworks to describe technological and innovation-driven systems.
        
        While the TAP equations describe the evolution of a (temporal) discrete system, most real world situations are characterised by a continuous time. Therefore we need to derive the continuous limit of \eqref{eq:TAP} in order to apply the TAP equations to a continuous scenario.
        For the passage to the continuum, we pose $t \rightarrow t_k=\frac{k}{n}T=\delta k\ $ for $\ T \gg 1$ and $\delta=T/n$ so that $\lim_{n \rightarrow \infty, T \rightarrow \infty} \delta = 0$. In addition, we pose: $\Sigma=\delta \sigma$ and $A_i = \delta \overline{\nu} \alpha_i$. For simplicity, we introduce the dimensional factor $\overline{\nu}$, which has the dimension of a rate, ensuring that the model parameters $\alpha_i$ are dimensionless.
        
        In this way, one has:
        \begin{equation*}
        M_{t_{k+1}} = (1-\Sigma) M_{t_k} + \sum_{i=1}^{M_{t_k}} A_i \binom{M_{t_k}}{i},
        \end{equation*}
        from which we get:
        \begin{equation*}
        \frac{M_{t_{k+1}}-M_{t_k}}{\delta} = -\frac{\Sigma}{\delta} M_{t_k} + \sum_{i=1}^{M_{t_k}} \frac{A_i}{\delta} \binom{M_{t_k}}{i},
        \end{equation*}
        and, in the limit $\delta \rightarrow 0$, one obtains the continuous limit of Eq.~(\eqref{eq:TAP}) as:
        \begin{equation}
        \frac{dM_t}{dt} = -\sigma M_{t} + \overline{\nu} \sum_{i=1}^{M_{t}} \alpha_i \binom{M_{t}}{i},
        \label{eq:TAP_continuous}
        \end{equation}
        Note that $\sigma$ has the dimension of a rate, while the $\alpha_i$ are dimensionless due to the explicit introduction of the dimensional factor $\overline{\nu}$. 
                
        Previous works already noticed that in the continuous limit, Eq.~(\eqref{eq:TAP}) leads to divergences in finite times~\cite{cortes2022tap}. Despite this, the direct relation between the TAP equations, hyperbolic functions and the emergence of a singularity has been previously overlooked. Indeed, the continuous limit of the TAP equations naturally leads to hyperbolic expressions. For instance, we can consider the simplest case, with no extinction rate ($\Sigma = 0$) and uniform coefficients ($A_i = A$ for any $i$). In this way \eqref{eq:TAP} simplifies to:
        \begin{equation}
        M_{t+1} = M_t + A \sum_{i=1}^{M_t} \binom{M_t}{i}
        \end{equation}
        Using Pascal's Triangle to evaluate the combinatorial term and taking the limit of continuous time, we obtain, analogously to \eqref{eq:TAP_continuous}:
        \begin{equation}
        \frac{dM_t}{dt} = \overline{\nu} \alpha (2^{M_t} - 1) \approx \overline{\nu} \alpha 2^{M_t}.
        \end{equation}
        The solution to this differential equation is the following function with a logarithmic divergence:
        \begin{equation}
        M_t = \log_2 \left( \dfrac{1}{2^{-M_0} -  \overline{\nu} \alpha \ln 2 (t-t_0)} \right) 
        \label{eq:sol_log2}
        \end{equation}
        where $M_0$ is the initial number of elements in the system at time $t_0$. The singularity of such equation lies at time $t^* = t_0 + (\overline{\nu} \alpha \ln 2)^{-1} 2^{-M_0}$. This result illustrates how singular trends observed in real-world data can arise from combinatorial models.
        
        Before diving into the detailed analysis of real-world systems, it is worth providing some remarks about the concept of singularity and its definition. Mathematically speaking, a function $f(t)$ has a singularity in $t^*$ if $t^*$ is an accumulation point of $f$. This occurs when the function diverges as $t$ approaches a value $t^* < \infty$:
        \[
         \lim_{t \to t^*} f(t) = \infty, \qquad t^* < \infty.
        \]
        It is important to stress that discrete models, despite a clear explosive behaviour, only have an accumulation point in $\infty$. This means that $f(t)$ can diverge only as $t \to \infty$. In other words, while we can define the discrete model also for $t>t^*$, i.e., we can compute $M_t$ for any $t<\infty$, in the continuous approximation the function $M_t$ does not exist for times larger than the singularity, so for $t>t^*$. Despite this limitation, hyperbolic functions have been widely used to interpret various systems, such as world population~\cite{korotayev2006world} and biological phase shifts in the field of biocosmology~\cite{korotayev2020twenty}.

        \begin{figure*}[t!]
            \includegraphics[width=\linewidth]{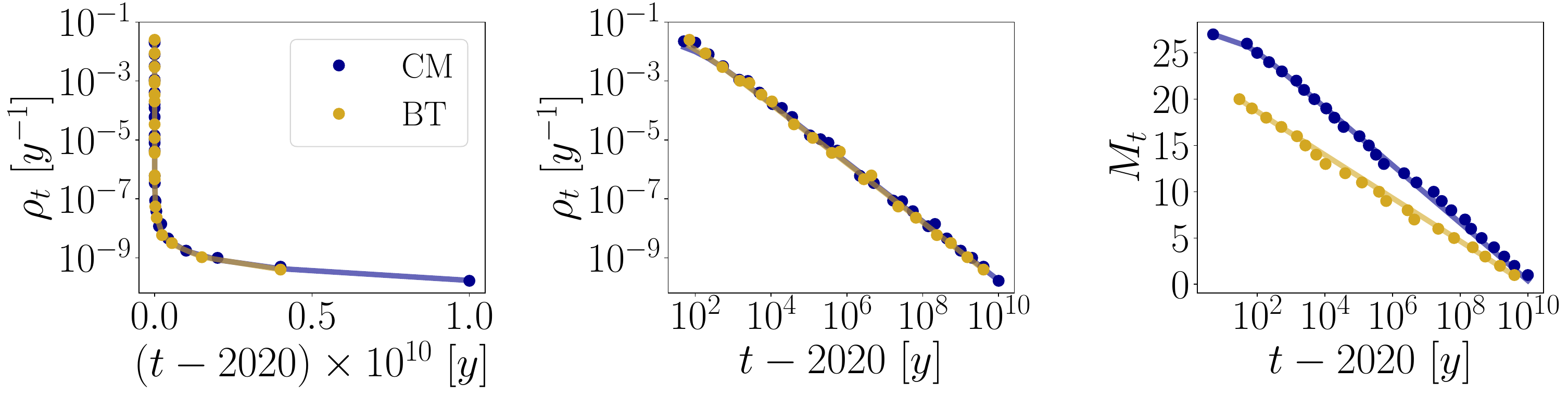}
            \caption{\textbf{Evolution of Canonical Milestones (CM) and Biological Phase Transitions (BT) over time.} 
            \textbf{Left.} Macrodevelopment rate $\rho_t$ as a function of time, computed as the inverse of the inter-event time between successive milestones. The log-linear scale highlights the hyperbolic growth pattern, as exponential growth would appear as a straight line on this scale. Solid lines show the results of the fit discussed in Appendix~\ref{app:numerical}. Time is expressed as the number of years before 2020 CE. \textbf{Center.} The same plot on a log-log scale, where the hyperbolic growth manifests as a straight line. The power-law exponent is $-1$, indicating a relationship of the form $y = 1/x$. \textbf{Right.} Number of events $M_t$ as a function of time, plotted on a linear-log scale to emphasize the logarithmic behaviour of milestone accumulation over time. The solid line represents the fit from the model in Eq.~\eqref{eq:model_bio}.}
            \label{fig:fig1}
        \end{figure*}

    \section{MODELING GROWTH AND INNOVATION}
            
        \label{sec:modeling}
            
        In this section, we conduct a comprehensive analysis of various scenarios, applying the concepts discussed earlier. We examine five distinct datasets: the number of Canonical Milestones and Biological Phase Transitions (Section~\ref{sec:biocosmology}), the growth of the world population and GDP (Section~\ref{sec:population_GDP}), and technological innovations from US Patents (Section~\ref{sec:US_patents}). For each case, we select a specific combinatorial model represented by a particular form of the TAP Equation, solve them, and demonstrate that they align with the results found in previous research. Finally, we validate the model results using real-world data.
            
        \subsection{Biological Phase Transitions}
            \label{sec:biocosmology}

            \begin{figure*}[t!]
                \includegraphics[width = \linewidth]{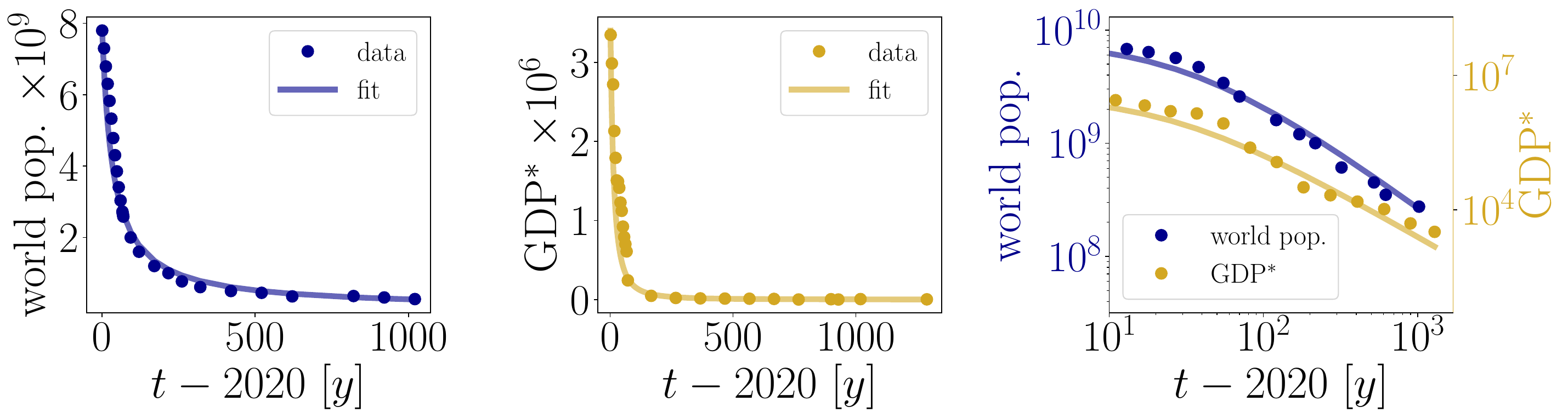}
                \caption{\textbf{World population and GDP growth over time.} 
                \textbf{Left.} World population size from 1000 to 2020 CE. The sudden explosion in the last century suggests a rate that exceeds exponential growth. The solid line represents the hyperbolic fit, as discussed in Appendix~\ref{app:numerical}. Time is expressed as the number of years before 2020 CE.
                \textbf{Center.} GDP aggregated for all countries, in billions of USD dollars (indicated as GDP$^*$ and adjusted for Parity for Purchase Power~\cite{bolt2020maddison}), from 700 to 2018 CE. The hyperbolic growth pattern is again evident from the sharp rise in GDP in recent years. The solid line shows the hyperbolic fit, as detailed in Appendix~\ref{app:numerical}. 
                \textbf{Right} World population and GDP evolution on a log-log scale. This scale highlights the hyperbolic growth, with a power law exponent of $1$ for world population and $2$ for GDP.}
                \label{fig:fig2}
            \end{figure*}
            
            In the realm of biocosmology, numerous studies have explored the emergence of paradigmatic events that have significantly influenced the history of the Universe and Life. These events include the origin of the Milky Way, the advent of life, the Cambrian explosion, the Hominoid revolution, and the emergence of democracy. Some studies refer to them as Canonical Milestones~\cite{kurzweil2004law, kurzweil2005singularity, modis2003limits}, while others as Biological Phase Transitions~\cite{panov2004avtomodel, panov2020twenty}. Despite slight variations in data representation, both terms refer to the same phenomena with similar meanings and intentions. The two corresponding time series are presented in Appendix~\ref{app:data}. 
            
            When examining the time series for the occurrence of these milestones, authors from both perspectives have recognised a hyperbolic growth behaviour (see Figure~\ref{fig:fig1}). Instead of analysing the direct count of milestones, most researchers consider the "inter time"~\cite{panov2004avtomodel, korotayev2020twenty}, which measures the time between two consecutive milestones. The inverse of the inter time, which represents the macrodevelopment rate $\rho_t$ and is effectively an approximation of the derivative of the milestone count $dM_t/dt$, closely follows a hyperbolic function with an exponent of $-1$:
            \[
            CM \ : \ \rho_t = \frac{2.054}{2029 - t}, \quad BT \ : \ \rho_t = \frac{1.886}{2027 - t}
            \]
            for Canonical Milestones (CM) and Biological Phase Transitions (BT), respectively. Here, time $t$ represents the number of years in the Common Era (CE), meaning that years before 0 CE (commonly referred to as BCE, Before Common Era) are represented as negative values.
            
            Starting with the general TAP equation~\eqref{eq:TAP} and assuming no extinction rate ($\mu = 0$), our choice of the model relies on the selection of the parameter form for $\alpha$. Given that we know no specific mechanisms leading to the emergence of these phase transitions, we assume that any possible recombination of existing elements could be significant. To account for the increasing difficulty of combining a greater number of elements, we follow the idea presented in~\cite{koppl2018simple, cortes2022tap} and set the coefficients to scale as a power law, $\alpha_i = \alpha^i$ for $\alpha \in (0,1)$. With this choice, the continuous TAP equation can be reformulated as~\cite{cortes2022tap}:
            \begin{equation}
            \frac{d M_t }{dt} = \overline{\nu} \sum_{i=1}^{M_t} \alpha^i \binom{M_t}{i} \approx \overline{\nu} (1+\alpha)^{M_t} = \overline{\nu} e^{\beta M_t}
            \label{eq:model_bio}
            \end{equation}
            with $\beta = \ln(1+\alpha)$, and $\overline{\nu}$ is the standard dimensional factor accounting for the time scale~\cite{koppl2018simple, cortes2022tap}. Its value can be directly inferred from the data, as explained in Appendix~\ref{app:numerical}. 
            
            From this expression, we derive the evolution of the macrodevelopment rate $\rho_t = \frac{dM_t}{dt}$:
            \begin{equation}
                \dfrac{d\rho_t}{dt} = \dfrac{d}{dt} \left( \dfrac{dM_t}{dt} \right) = \beta \dfrac{dM_t}{dt} \overline{\nu} e^{\beta M_t} \approx \beta \rho_t^2
                \label{eq:model_rho}
            \end{equation}
            which matches the dynamics proposed in previous studies~\cite{korotayev202021st}. The scaling factor $\overline{\nu}$ disappears in the equation for the rate, as any rescaling of time proportionally affects both the rate and the time variable. The solution of this differential equation is a hyperbolic function with an exponent of $-1$:
            \begin{equation}
                \rho_t \approx \dfrac{1}{\rho_0^{-1} - \beta(t-t_0)}
                \label{eq:derivative_tap_bio}
            \end{equation}
            The explicit expression for the number of milestones $M_t$ can be obtained from equation~\eqref{eq:model_bio} (details are provided in Appendix~\ref{app:general}):
            \begin{equation}
                M_t \approx \frac{1}{\beta } \ln \left( \dfrac{1}{e^{-\beta M_0} - \overline{\nu} \beta(t-t_0)}\right)
            \label{eq:solution_tap_bio}
            \end{equation}

            This model replicates the same analytical behaviour observed in previous analyses~\cite{panov2004avtomodel, kurzweil2005singularity}. Importantly, these expressions emerge naturally from the combinatorial model's structure, offering a straightforward physical interpretation rather than being derived solely from observed time series data. We can fit the time series using our model and compute the coefficient $\alpha$ that reproduces the actual number of milestones $M_t$. The results of this numerical analysis are presented in Figure~\ref{fig:fig1} and in Appendix~\ref{app:numerical}. Following a convention widely adopted in the literature~\cite{korotayev2020twenty}, the time axis is shifted to represent the number of years before the present. Specifically, we use the horizontal axis $\tau = 2020 - t$, where $2020$ CE is taken as the reference year.

            The values of the coefficients in the two cases are $\alpha_{CM} =  0.670$ and $\alpha_{BT} = 0.837$. These model parameters represent the fraction of combinations occurring at any given time $t$, which indicates the rate of successful combinations that actually take place at each time step. We can refer to $\alpha$ as the macrodevelopment parameter of the system. The small difference in their values for Canonical Milestones and Biological Phase Transitions is due to the different events considered in the two time-series. However, the two parameters are comparable, suggesting a similar underlying process. It is also important to note that these parameter values cannot be used for long-term predictions of the time series. This limitation arises from the increasing divergence between the continuous model and the discrete one as the system approaches the singularity.
            
            Additionally, the expressions we derived present a singularity at $t = t_0 + \beta^{-1} \rho_0^{-1}$, a time value at which the function theoretically diverges. However, the discrete version of the model in equation~\eqref{eq:model_bio} does not exhibit any divergence in finite time, as it happens for real-world systems. We will explore the significance of this concept in Section~\ref{singularity}. 

        \subsection{World Population and GDP}
            \label{sec:population_GDP}

            \begin{figure*}[t!]
                \includegraphics[width=1\linewidth]{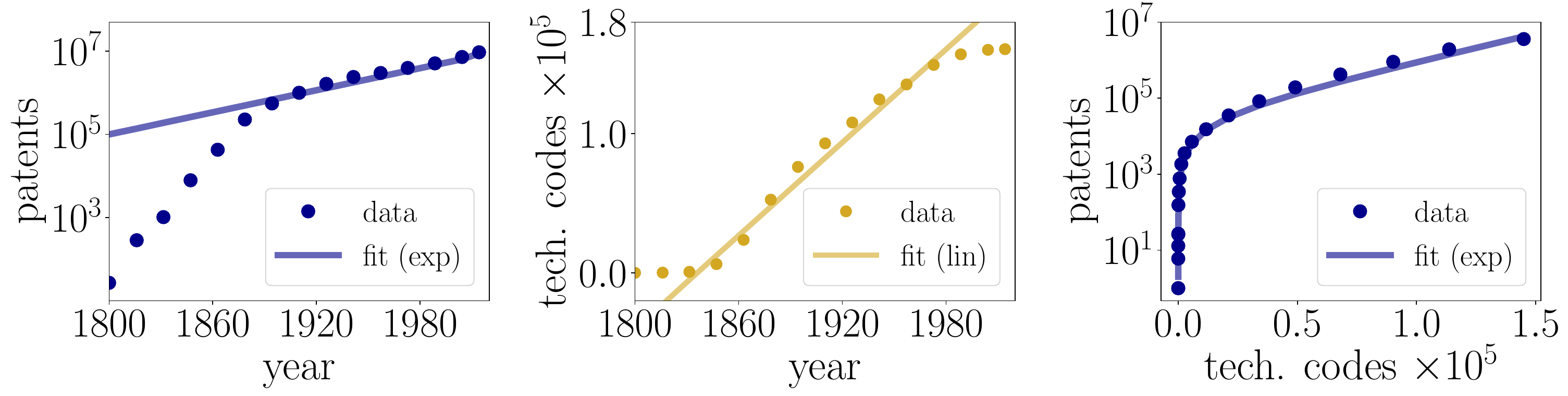}
                \caption{\textbf{Evolution of the number of patents and technological codes over time.}  
                \textbf{Left.} Number of patents over time, shown on a log-linear scale. There is a rapid growth phase until approximately 1860, after which the trend becomes exponential. The solid line represents an exponential fit of the data starting from 1900. Time represents the number of years in the Common Era (CE). \textbf{Center.} Number of distinct technological codes as function of time on a linear scale. The solid line, following the empirical trend, highlights the linear growth of technological codes observed up to $1980$. \textbf{Right.} Number of patents as a function of technological codes, presented on a log-linear scale, which highlights the exponential growth of patents. As discussed in Appendix~\ref{app:numerical}, the solid line represents the fit.}
                \label{fig:US_patents}
            \end{figure*}
            
            The growth of world population size over the years surpasses exponential rates, as widely recognised in prior literature~\cite{kremer1993population, kapitza1996phenomenological}. The seminal paper from 1960, "Doomsday: Friday, 13 November, A.D. 2026"~\cite{von1960doomsday}, was among the first to identify this phenomenon. The authors observed that demographic growth from the year 1 CE to 1958 CE followed a hyperbolic pattern:
            \begin{equation}
                N_t = \dfrac{C}{t^* - t}
                \label{eq:world_pop}
            \end{equation}
        with approximate constants $C \approx 2.15 \cdot 10^5 \ [y]$ and $t^* \approx 2026 \ [y]$. The value $t^*$ represents the mathematical singularity of the hyperbolic curve, which was used to speculate about future population growth. This hyperbolic pattern has been verified from earlier periods~\cite{kremer1993population, kapitza1996phenomenological} up to a million years ago and continues to be relevant for some years beyond 1958~\cite{korotayev2020twenty}. 
            
            Typically, these datasets are interpreted using dynamic models that consider the tradeoff between mortality and fecundity~\cite{lee1980aiming} without explicit accounting for individual interaction. In contrast, the combinatorial approach based on TAP has been employed to explain world population growth~\cite{koppl2018simple}, although it does not focus on the specific types of interactions involved. Here, we show that by incorporating particular interaction terms into the combinatorial model, we can reproduce and predict the observed hyperbolic growth. 
            
            Starting with the general TAP equation~\eqref{eq:TAP}, we select appropriate values for the parameters $\alpha_i$. Demographic growth is driven by reproductive processes, which can be modelled as pair interactions.  We then hypothesize that the driving term is for $i=2$, corresponding to the number of pairs in the system:
            \begin{equation}
                \dfrac{dN_t}{dt} = \overline{\nu} \alpha^2 \binom{N_t}{2} = \overline{\nu} \alpha^2 \dfrac{N_t(N_t-1)}{2} \approx \overline{\nu} \gamma N_t^2 = \nu_{\gamma} N_t^2
                \label{eq:world}
            \end{equation}
            Given $N_t \gg 1$, we can neglect the linear term, focusing on the quadratic term and redefining $\gamma = \alpha^2/2$ as a birth rate. Notice that we do not consider direct mortality effects, so $\mu=0$ in Eq.~\eqref{eq:TAP}. Additionally, the usual dimensional factor $\overline{\nu}$ is reabsorbed into the parameter  $\nu_{\gamma} = \overline{\nu} \gamma$, which now carries the dimension of a rate.         
            
            The solution to equation~\eqref{eq:world} produces the hyperbolic curve (details are reported in Appendix~\ref{app:general}):
            \begin{equation}
                N_t \approx \dfrac{1}{N_0^{-1} - \nu_{\gamma} (t - t_0)}
                \label{eq:world_sol}
            \end{equation}
            This result aligns with previous analyses~\cite{korotayev2020twenty}. It is important to note that this solution is obtained using the continuous approximation of the model, while a discrete model would not exhibit any singularity. Results of the fit of the macrodevelopment parameter $\gamma$ are reported in Figure~\ref{fig:fig2} and in Appendix~\ref{app:numerical}. As in the previous case, the time scale is shifted by convention to represent the number of years before $2020$ CE.
            
            Furthermore, some works~\cite{korotayev202021st, korotayev2020twenty} mistakenly equate the hyperbolic patterns of demographic growth with those of Canonical Milestones, suggesting they are generated by the same process. However, these situations are distinct, as demonstrated by the different models used to fit the data. For Canonical Milestones ($M_t$), the hyperbolic pattern pertains to the rate, described by the equation~\eqref{eq:derivative_tap_bio}, which involves all combinatorial terms. In contrast, for the world population, the hyperbolic pattern emerges directly in $N_t$ (Eq.~\eqref{eq:world_sol}), where only pair interactions are considered.
            
            Finally, we extend our analysis to GDP per capita, reaffirming previous findings within the new combinatorial framework. Many researchers have postulated a relationship between GDP growth and world population~\cite{malthus2023essay, acemoglu2008introduction}. According to Malthusian macroeconomic principles, the GDP $K_t$ grows at a rate proportional to the size of the population $N_t$. We can represent this growth rate with a combinatorial term proportional to $K_t N_t$~\cite{korotayev2009compact}. Additionally, empirical data show that GDP's behaviour over the years closely mirrors the square of the population $N_t \approx \sqrt{K_t}$~\cite{korotayev2005compact}. Using these observations, the GDP growth rate can be expressed as:
            \begin{equation}
                \dfrac{dK_t}{dt} = \overline{\nu} \delta K_t N_t \approx \nu_{\delta} K_t^{3/2}
            \end{equation}
            where $\nu_{\delta}$ is the macrodevelopment parameter, which has the dimension of a rate. The solution is a hyperbolic curve with an exponent of 2 (analytical details are given in Appendix~\ref{app:general}):
            \begin{equation}
                K_t \approx \dfrac{1}{\left(K_0^{-1/2} - \nu_{\delta} (t - t_0) \right)^2}
                \label{eq:GDP}
            \end{equation}
            This result matches the expression fitted to the data in previous works~\cite{korotayev2009compact}. Once again, this hyperbolic behaviour is explained within the combinatorial growth framework. Results of the fit of the parameter $\delta$ are reported in Figure~\ref{fig:fig2} and Appendix~\ref{app:numerical}. 

        \subsection{Technological Innovations: the Case of US Patents}
        \label{sec:US_patents}

            Technological innovations, as demonstrated through patenting activities, provide an effective example of combinatorial growth. Patents are characterised by a set of codes, which reveal the underlying technological capabilities used in the invention. As a result, patents naturally emerge as combinations of technological codes. However, this system differs significantly from those previously explored, as the number of patents displays an exponential, rather than hyperbolic, increase over time~\cite{youn2015invention}.

            Due to this exponential behaviour, previous frameworks have approached US patents differently~\cite{youn2015invention}. Specifically, the TAP equation has been used to model the temporal evolution of patents, as well as other aspects such as the distribution of descendants~\cite{sole2016singularities, koppl2023explaining}. However, the absence of a singularity in the case of US patents requires introducing an additional aging process into the model. This process slows down the hyperbolic growth, leading to the emergence of an exponential behaviour.
            Here, we demonstrate that our framework captures the dynamics of patents, reproducing the essential features of the data directly from the TAP equation.

            The left panel of Figure~\ref{fig:US_patents} illustrates the exponential growth of the number of patents over time when plotted on a log-linear scale. Conversely, the behaviour of technological codes is linear rather than exponential. As shown in the right panel of Figure~\ref{fig:US_patents}, plotting the number of patents relative to the number of distinct technological codes reveals an exponential relationship. 

            Notice that, unlike previous cases, here patents and technological codes do not share the same nature. They represent, indeed, two different conceptual categories. For this reason, we exploit the idea that patents are created through the combination of technological codes, and we model the evolution of patents with respect to these codes. Patents can theoretically involve any number of codes, with larger combinations presenting more challenges to be realised. Therefore, we consider all terms within equation~\eqref{eq:TAP} to establish an exponential model, similar to the approach outlined in Section~\ref{sec:biocosmology}:
            \begin{equation}
            \frac{dP_C}{dC} = \sum_{i=1}^{C} \alpha^i \binom{C}{i} \approx e^{\beta C}
            \label{eq:patent}
            \end{equation}
            with $\beta = \ln(1+\alpha)$, where $\alpha$ is the usual macrodevelopment rate of the system. For simplicity, we treat all quantities (patents and technological codes) as dimensionless, as time is not explicitly present in this model. As a consequence, in this case, there is no need to explicitly consider any dimensional factor. Here, $C$ represents the number of technological codes, while $P_C$ denotes the number of patents when the number of technological codes is $C$. In this model, we treat the number of technological codes as a time scale $C$. Consequently, the solution of equation~\eqref{eq:patent} is as follows:
            \begin{equation}
            P_C \propto \dfrac{1}{\beta} e^{\beta C}, \quad C_P \propto \dfrac{1}{\beta} \ln \beta P
            \label{eq:patent_sol}
            \end{equation}
            This equation replicates the trends observed for patents and technological codes, as shown in the right panel of Figure~\ref{fig:US_patents}. We also present the specific curve obtained by fitting the data. More details on the coefficients used in this model are provided in Appendix~\ref{app:numerical}. 
            
Moreover, the relationships described in Eq.~\eqref{eq:patent_sol} enable us to recover the temporal growth of the number of patents, as shown in the left panel of Figure~\ref{fig:US_patents}. In regions where $C(t)$ approximates linearity, an exponential growth pattern emerges for patents such that $P(t) \propto e^t$. It should be noted, however, that the exponential relationship between the two quantities described by equation~\eqref{eq:patent_sol} remains valid regardless of the specific temporal behaviour of $C(t)$ or $P(t)$. It is important to stress that, in this case, the model does not produce any singularity, even if the underlying mechanism is analogous to that applied to other systems. This showcases the versatility of the TAP equation and the theory of combinatorial innovation to describe a wide variety of different systems with similar but yet different underlying dynamics.

\section{Discussion}
\label{singularity}

The mathematical singularity arising from the hyperbolic behaviours described in Section~\ref{sec:modeling} has been subject to various interpretations. The concept of ``Singularity'' was already mentioned in the work of Von Foerster et al.~\cite{von1960doomsday}, who provocatively identified a specific date for the so-called "Doomsday," projected to occur on 13 November 2026. More recently, scholars offered two main interpretations of this concept. The first, as suggested by~\cite{von1960doomsday}, argues that such a singularity indicates that rapid growth is unsustainable in real finite systems. As discussed in~\cite{koppl2023explaining}, a singularity "would be a regime change. It would mark the point at which our model no longer applies”, highlighting its conceptual importance as a transition point, beyond which standard modelling may fail. This view implies that a slowdown in growth must occur before any potential explosive scenario~\cite{korotayev2020twenty, korotayev202021st, modis2013singularity}. In this interpretation, the mathematical divergence can be seen as a potential indicator of this deceleration getting closer. In fact, in many systems, it is reasonable to expect a deceleration due to the presence of limiting factors such as the finite availability of resources~\cite{villani2023super}. However, our mathematical analysis reveals that inferring this slowdown from the data may be very tricky. First, the discrete versions of the models, which typically represent the actual dynamics of real-world systems, do not exhibit any singularity. The discrete models can be indeed used to forecast growth beyond this theoretical limit, as detailed in Appendix~\ref{app:numerical}. Secondly, our models do not explicitly account for external limiting factors, which are the true drivers of any changes in the system. As a result, these simple models are not able to predict any shift in the system's evolution. This makes the singularity value practically unrelated to any constraints or deceleration.
        
Conversely, some authors, particularly in the Big History field, view the singularity as an indicator of upcoming transitions or technological shifts~\cite{kurzweil2005singularity, panov2020singularity, nazaretyan2015megahistory}. For instance, the concept of Threshold 9 in the field of Big History~\cite{nazaretyan2016non} suggests that humanity is approaching a profound global paradigm shift (the threshold) that will significantly impact both technology and evolution~\cite{kurzweil2005singularity}. Koppl et al.~\cite{koppl2023explaining} note, however, that the nature of a technological singularity remains uncertain—it "could be heaven, hell, or something else altogether." Other authors even speculate about post-singularity civilisations~\cite{panov2011post}. This second interpretation often relies on extrapolations from data without direct evidence that current growth rates will continue indefinitely, positioning such analyses more within the realm of philosophical speculation, though they inspire fascinating and stimulating discussions. From a technical perspective, attempts to extrapolate precise years from the data can lead to peculiar results. For example, in~\cite{von1960doomsday}, the authors calculated the value $N_t = 1$ from the world population curve and interpreted it as "Adam's birth," resulting in a date nearly 200 billion years in the past. Similarly, some predictions appear strikingly accurate but are essentially coincidental, such as the singularity in GDP per capita growth estimated for 2005~\cite{korotayev2009compact}, which is closely aligned with the 2008 financial crisis. 

Finally, it is important to stress that estimating a reliable singularity value is almost impossible due to the high value of statistical error. As noted in~\cite{von1960doomsday}, the year obtained from fitting data can vary substantially depending on the choice of the last observation. For instance, if Charlemagne, with data until 800 CE, had predicted a singularity, he would have projected it 300 years into the future, while Napoleon, using more recent data, would have predicted it just 30 years ahead. In the very same way, our estimates of singularities, presented in Appendix~\ref{app:numerical}, differ from previous analyses in the literature due to the updated data. In certain cases, large time intervals considered in the analysis result in wide confidence bounds for the numerical value of the singularity, making it even more difficult to obtain a reliable estimate.

\section{Conclusions}
    
This study investigates the dynamics of growth and innovation, offering a unified approach for modelling and understanding it. Innovation dynamics often exhibit hyperbolic trends, which have been studied and interpreted in various ways in the literature. However, a coherent framework that models these processes from a physical and mathematical perspective has been lacking. By leveraging a combinatorial approach based on the Theory of Adjacent Possible (TAP), we provide a comprehensive theoretical foundation to interpret these phenomena. We investigated several distinct scenarios, demonstrating how our framework can successfully replicate the main characteristics of growth trends. We analysed four key cases: biological milestones (Canonical Milestones and Biological Phase Transitions), world population, GDP growth, and technological innovations as seen through U.S. patent data. In each case, the models allow us to estimate the macrodevelopment parameter from the data, enabling us to quantify the system's growth rate. 

This paper demonstrates how a combinatorial framework can offer a coherent interpretation of growth and innovation processes across diverse domains. Future research should explore applying these models to other complex systems, such as biological or evolutionary processes, where similar growth dynamics are observed. The main limitation of our approach is that it does not consider any external factor acting on the system, which is relevant in most cases. Given the simplicity and interpretability of our approach, future work could focus on incorporating external constraints, such as resource limitations, to produce more accurate descriptions and forecasts of system behaviour. Another promising direction is investigating the connection between this framework and the triggering mechanisms described in other discovery processes~\cite{tria2014dynamics, di2025dynamics, bellina2024time}. In this view, the emergence of innovations is explained through exploring a network, where each novelty opens up the possibility for further developments. Establishing a connection between these frameworks could further advance our understanding of innovation processes and achieve a comprehensive theory of innovation.

\appendix

\section{Data sources}
    \label{app:data} 
    \subsection*{Canonical Milestones}

        Modis-Kurzweil time series~\cite{modis2003limits, kurzweil2005singularity} considers key evolutionary and technological milestones, focusing on the interplay between demographic growth and technological innovation. They are also referred to as \textit{Canonical Milestones}. Starting from the Big Bang, it outlines 27 major critical events in the history of the universe, from the formation of galaxies to the rise of the Internet. The time in parentheses refers to the number of years before 2020.
    
        \small{
            \begin{enumerate}
            \item \textbf{Origin of Milky Way and First Stars (10 billion years)}: The formation of the Milky Way galaxy and the first stars.
            \item \textbf{Origin of Life on Earth, Formation of the Solar System (4 billion years ago)}: The emergence of life on Earth and the formation of the solar system and Earth's oldest rocks.
            \item \textbf{First Eukaryotes and Invention of Sex (2 billion years)}: The development of eukaryotic cells, the emergence of sexual reproduction, atmospheric oxygen, photosynthesis, and plate tectonics.
            \item \textbf{First Multicellular Life (1 billion years)}: The appearance of multicellular organisms such as sponges, seaweeds, and protozoans.
            \item \textbf{Cambrian Explosion and First Vertebrates (430 million years)}: A period of rapid diversification of life, with the emergence of invertebrates, vertebrates, plants, and amphibians.
            \item \textbf{First Mammals and Dinosaurs (210 million years)}: The rise of mammals, birds, and dinosaurs.
            \item \textbf{First Flowering Plants (139 million years)}: The appearance of angiosperms, or flowering plants.
            \item \textbf{Mass Extinction and First Primates (54.6 million years)}: An asteroid impact causes mass extinction, including the dinosaurs, and the emergence of primates.
            \item \textbf{First Hominids (28.5 million years)}: The appearance of early human ancestors.
            \item \textbf{First Orangutans and Proconsul (16.5 million years)}: The origin of orangutans and the early ape genus Proconsul.
            \item \textbf{Chimpanzees and Humans Diverge (5.1 million years)}: The evolutionary split between humans and chimpanzees, with the earliest evidence of bipedalism in hominids.
            \item \textbf{First Stone Tools and Homo Erectus (2.2 million years)}: The development of early stone tools and the appearance of Homo erectus.
            \item \textbf{Emergence of Homo Sapiens (555,000 years ago)}: The first appearance of anatomically modern humans.
            \item \textbf{Domestication of Fire (325,000 years)}: Homo heidelbergensis begins controlling fire for cooking and warmth.
            \item \textbf{Human DNA Differentiation (200,000 years)}: The divergence of different human DNA types.
            \item \textbf{Emergence of Modern Humans (105,700 years)}: The earliest evidence of modern humans and their burial practices.
            \item \textbf{Rock Art and Protowriting (35,800 years)}: The creation of early forms of symbolic communication through rock art.
            \item \textbf{Techniques for Starting Fire (19,200 years)}: The development of techniques for creating fire independently.
            \item \textbf{Invention of Agriculture (11,000 years)}: The shift from hunter-gatherer societies to settled farming communities.
            \item \textbf{Invention of the Wheel and Writing (4907 years)}: The development of the wheel, writing systems, and the rise of early civilizations in Egypt and Mesopotamia.
            \item \textbf{Democracy and the Axial Age (2437 years)}: The rise of city-states, democracy in Greece, and the teachings of figures like Buddha.
            \item \textbf{Invention of Zero and Fall of Rome (1440 years)}: The development of the concept of zero and the fall of the Roman Empire.
            \item \textbf{Renaissance and the Scientific Method (539 years)}: A period of cultural rebirth in Europe, marked by the discovery of the New World and the development of the scientific method.
            \item \textbf{Industrial Revolution (225 years)}: The advent of steam engines, political revolutions, and the rise of industrial economies.
            \item \textbf{Modern Physics and Technological Advances (100 years)}: Major breakthroughs in physics, radio, electricity, automobiles, and aviation.
            \item \textbf{Nuclear Energy and the Cold War (50 years)}: The discovery of DNA's structure, the invention of the transistor, and the onset of the Cold War.
            \item \textbf{Internet and Human Genome Sequencing (5 years)}: The sequencing of the human genome and the rise of the internet as a transformative technology.
        \end{enumerate}
        }

\normalsize

    \subsection*{Biological Phase Transitions}

        Panov time series~\cite{panov2020singularity, korotayev202021st} represents a historical trajectory of complexity growth, marking key transitions in the development of life and civilization on Earth. They are also referred to as \textit{Biological Phase Transitions}. This series of 20 events covers critical points in universal and human history, from the origin of life to modern history. Each event signifies a substantial increase in complexity, accelerating over time as each phase becomes shorter. The time in parentheses refers to the number of years before 2020.
    
        \small{
        \begin{enumerate}
            \item \textbf{Origin of Life (4 billion years)}: The emergence of life, with the biosphere dominated by nucleus-less prokaryotes for the first 2-2.5 billion years.
            \item \textbf{Neoproterozoic Revolution (1.5 billion years)}: Cyanobacteria enrich the atmosphere with oxygen, leading to the extinction of many anaerobic prokaryotes and the rise of aerobic eukaryotes and multicellular life.
            \item \textbf{Cambrian Explosion (590-510 million years)}: A rapid diversification of life forms, leading to the emergence of all modern animal phyla, including vertebrates.
            \item \textbf{Reptiles Revolution (235 million years)}: A mass extinction wipes out most Paleozoic amphibians, allowing reptiles to dominate terrestrial life.
            \item \textbf{Mammalian Revolution (66 million years)}: The extinction of the dinosaurs marks the beginning of the dominance of mammals on land.
            \item \textbf{Hominoid Revolution (25-20 million years)}: A significant evolutionary event that leads to the rise of numerous Hominoidae genera, many more than exist today.
            \item \textbf{Quaternary Period (4.4 million years)}: The first primitive members of the Homo genus diverge from other Hominoidae.
            \item \textbf{Palaeolithic Revolution (2.0-1.6 million years)}: The emergence of Homo habilis and the first use of stone tools.
            \item \textbf{Chelles Period (700,000-600,000 years)}: The discovery of fire and the rise of Homo erectus.
            \item \textbf{Acheulean Period (400,000 years)}: The standardization of symmetric stone tools.
            \item \textbf{Neanderthal Culture Revolution (150,000-100,000)}: Homo sapiens neanderthalensis appears with fine stone tools and burial practices, suggesting the beginnings of primitive religion.
            \item \textbf{Upper Palaeolithic Revolution (40,000 years)}: Homo sapiens sapiens becomes the dominant species, developing advanced hunting tools and imitative art.
            \item \textbf{Neolithic Revolution (12,000-9,000 years)}: The shift from foraging to food production, marking the rise of agriculture.
            \item \textbf{Urban Revolution (6000-5000 years)}: The emergence of state formations, written language, and the first legal codes.
            \item \textbf{Imperial Antiquity and Iron Age (2800-2500 years)}: The rise of empires and the Axial Age, marked by cultural revolutions and thinkers like Zarathustra, Socrates, and Buddha.
            \item \textbf{Beginning of the Middle Ages (1600-1400 years)}: The fall of the Western Roman Empire, the spread of Christianity and Islam, and the dominance of feudal economies.
            \item \textbf{Modern Period and First Industrial Revolution (570-470 years)}: The rise of manufacturing, the printing press, and the cultural revolutions of the early modern era.
            \item \textbf{Second Industrial Revolution (190-180 years)}: The mechanization of industry, with the rise of steam power, electricity, and early globalization through the telegraph.
            \item \textbf{Information Revolution (70 years)}: The transition to post-industrial society, where the majority of workers in industrialized countries are employed in information-related fields or services.
            \item \textbf{Crisis and Collapse of the Communist Block (30 years)}: Information globalization accelerates following the political and economic changes marked by the end of the Cold War.
        \end{enumerate}
        }

    \subsection*{World Population}
    
        The data on world population growth was retrieved from Gapminder (https://www.gapminder.org/). As the platform states, "Gapminder combines data from multiple sources into unique coherent time series that can’t be found elsewhere." The time series provides estimates of world population size from $10,000$ BCE to the present day. For this analysis, we considered the period from $1000$ to $2020$ CE.
    
    \subsection*{GDP}
    
        The data on GDP growth was obtained from the Maddison Project Database 2020~\cite{bolt2020maddison}. According to the website, "The Maddison Project Database provides information on comparative economic growth and income levels over the very long run. The 2020 version of this database covers 169 countries and the period up to 2018". We aggregated the time series of all $169$ countries to create a global GDP time series, considering data from $700$ CE to $2018$ CE.
    
    \subsection*{US Patents}
    
        The data on U.S. patents was retrieved from the United States Patent and Trademark Office (USPTO) (http://www.uspto.gov/patents/)~\cite{lyons2003united}. This source provides time series data on all U.S. patents from $1800$ CE to $2020$ CE. Patents are classified by technological codes according to the U.S. Patent Classification System, where each code is represented by an alphanumeric string. From this data, we constructed a time series for both the number of patents and the associated technological codes over time.

\section{General Solution of Hyperbolic Differential Equation}
\label{app:general}

    In this appendix, we explicitly solve the differential equation governing the models presented in Section~\ref{sec:modeling}. Throughout the following derivation, time is expressed in years $[y]$.
    
    First, consider the differential equation of the form displayed in Section~\ref{sec:biocosmology}, given by
    \[
        \dfrac{dx}{dt} = a e^{b x},
    \]
    which can be solved by the separation of variables:
    \[
        e^{-b x} dx = a dt.
    \]
    Integrating $dx$ from $x_0$ to $x_t$, and $dt$ from $t_0$ to $t$, we obtain:
    \[
        -\dfrac{e^{-b x_t}}{b} + \dfrac{e^{-b x_0}}{b} = a(t - t_0).
    \]
    Inverting this relation, we get the explicit form of the solution:
    \[
        x_t = \dfrac{1}{b} \ln \left( \dfrac{1}{e^{-b x_0} - a b(t-t_0)}  \right).
    \]
    This expression presents a mathematical singularity at the point $t = t _0 + a^{-1} b^{-1} e^{-b x_0}$. Note that Eq.~\eqref{eq:sol_log2} is derived from:
    \[
        \dfrac{dM_t}{dt} \approx \overline{\nu} \alpha 2^{M_t} = \overline{\nu} \alpha e^{\ln 2 M_t}
    \]
    with $a = \overline{\nu} \alpha$ and $b = \ln 2$.
    
    For a differential equation of the form displayed in Section~\ref{sec:population_GDP}:
    \[
        \dfrac{dx}{dt} = a x^{b},
    \]
    we solve by separation of variables:
    \[
        x^{-b} dx = a dt.
    \]
    Integrating $dx$ from $x_0$ to $x_t$, and $dt$ from $t_0$ to $t$, we obtain:
    \[
         \dfrac{x^{1-b}}{1-b} - \dfrac{x_0^{1-b}}{1-b} = a(t - t_0).
    \]
    Inverting this relation, we obtain the explicit form of the solution:
    \[
        x_t = \left[ (1-b) \left( a (t - t_0) - \frac{x_0^{1-b}}{b - 1} \right)\right]^{\frac{1}{1-b}}.
    \]
    When $b > 1$, this expression presents a divergence at the point $t^* = t_0 + a^{-1} (b - 1)^{-1} x_0^{1-b}$.
    
    Finally, the differential equation reported in Section~\ref{sec:US_patents} presents simple exponential growth:
    \[
        \dfrac{dP}{dC} = e^{a C}.
    \]
    Integrating both sides, we get
    \[
        P_C = \dfrac{e^{a C}}{a} - \dfrac{e^{a C_0}}{a} + P_0.
    \]
    This expression shows no divergence in finite time, meaning there is no singularity.

\section{Results of numerical analysis}
\label{app:numerical}
    
    In this section, we present the results of fitting the models to real data and discuss the meaning of the parameters used. For all analyses, the time variable $t$ is defined as the time before 2020, i.e.,  $\tau = 2020 - t$, with the year 2020 corresponding to $\tau = 0$.
    
    For both Biological Phase Transitions and Canonical Milestones, the macrodevelopment rate $\rho_t = \frac{dM_t}{dt}$ follows the expression (Eq.~\eqref{eq:derivative_tap_bio}):
    \[
        \rho_t = \dfrac{1}{\rho_0^{-1} - \beta (t - t_0)}.
    \]
    With the time-shift $\tau = 2020 - t$, as also used in Figure~\ref{fig:fig1}, the macrodevelopment rate reads:
    \[
        \rho_{\tau} = \dfrac{1}{\rho_0^{-1} - \beta (2020 - \tau - t_0)} = \dfrac{1}{\rho_0^{-1} + \beta (\tau - \tau_0)},
    \]
    where $\tau_0 = 2020 - t_0$ is the shifted initial condition. This expression was used in the fit.
    
    We fitted $1/\rho_{\tau}$ against $\tau$ using a linear model, setting $\tau_0 = 50$ ($t_0 = 1970$) for Canonical Milestones and $\tau_0 = 70$ ($t_0 = 1950$) for Biological Phase Transitions. The initial values of $\rho_0$ were taken as $\rho_0^{CM} = 0.0222$ and $\rho_0^{BT} = 0.0250$ from the time series, respectively.
    
    We fitted the parameter $\beta$ and then calculated the macrodevelopment rate $\alpha$ using the relation $\beta = \ln(1 + \alpha)$:
    \[
    \begin{cases}
        \beta_{CM} = 0.513 \pm 0.007 \\
        \alpha_{CM} = 0.670 \pm 0.012
    \end{cases}
    \]
    \[
    \begin{cases}
        \beta_{BT} = 0.608 \pm 0.002 \\
        \alpha_{BT} = 0.837 \pm 0.004
    \end{cases}
    \]    
    The singularity time $\tau^*$ can be calculated as:
    \[
        \tau^* = \tau_0 - \frac{1}{\rho_0 \beta}
    \]
    which gives:
    \[
    \tau^*_{CM} = -37.7 \pm 17.8 \ [y], \ (t^*_{CM} \approx 2058 \ [y])
    \]
    \[
\tau^*_{BT} = 4.2 \pm 2.9 \ [y], \ (t^*_{BT} \approx 2016 \ [y])
    \]
    Note that, from our definition of $\tau$, its negative values correspond to times after 2020 CE, while positive values indicate times before 2020 CE. In the case of Biological Phase Transitions, the singularity appears in the past because the time series ends $70$ years before 2020. This highlights the ambiguity and arbitrariness in estimating the exact singularity time, which strongly depends on the chosen endpoint of the dataset. Our results also differ from those reported in~\cite{korotayev202021st}, probably due to different fitting techniques. Additionally, the quite large error margins prevent a precise estimate of the singularity year.
    
    The macrodevelopment parameter $\alpha$ represents the combinatorial success rate, quantifying the fraction of potential combinations that actually produce successful innovations. The two parameters $\alpha_{CM}$ and $\alpha_{BT}$ are close in magnitude, suggesting a similar underlying process, irrespective of the specific events considered in the two time series.

    Regarding the number of events $M_t$, the continuous dynamics described in equation~\eqref{eq:solution_tap_bio}, with parameter $\beta$ obtained by fitting the macrodevelopment rate, may introduce errors when the time intervals between events are large. This is particularly relevant in the case of Canonical Milestones and Biological Phase Transitions, where we have only $27$ or $20$ data points over a time range of $10^{10}$ years.  In Eq.~\eqref{eq:solution_tap_bio}, $M_t$ is computed by a continuous integration of $\rho_t$, which can lead to discrepancies compared to the actual discrete dynamics when the time intervals are large. This issue does not arise for world population, GDP, and patents, where denser data sets minimize errors in the transition from discrete to continuous models.
    
    To avoid these errors, the sequence of events $M_t$ can be reconstructed directly using the discrete model based on $\rho_t$:
    \begin{equation}
        \begin{cases}
            \rho_{t+\Delta t} = \rho_t + \beta \rho_t^2 \Delta t, \\
            M_{t + \Delta t} = M_t + \rho_t \Delta t.
        \end{cases}
        \label{eq:model_bio_discrete}
    \end{equation}
    Here, the first equation discretizes equation~\eqref{eq:model_rho}, while the second reproduces the sequence $M_t$ shown in Figure~\ref{fig:fig1}. This ensures that the dynamics remain faithful to the discrete model, particularly when the time intervals between events are large, as is the case for milestones.

    Using the discrete model, we can estimate the timing of new Canonical Milestones in the short-term future. It is important to note that the continuous version of the model provides an exact analytical continuation of the discrete one only far from the singularity. Parameters obtained by fitting the continuous model can reliably predict $\rho_t$ over short time intervals following the last observed event. To extend the prediction range, the sequence might be updated with new data points (either real or estimated), and the parameters should be adjusted based on the updated sequence.
    
    From the second equation in~\eqref{eq:model_bio_discrete}, we can compute the macrodevelopment rate $\rho_t$ at the next stage of the evolution and subsequently estimate the number of milestones $M_t$. Let's focus on Canonical Milestones. The inverse of $\rho_t$ represents the inter-time, i.e., the time interval between two consecutive events. Starting with $\rho_0 = 0.022$ at $\tau_0 = 50$, the next event $M_t = 27$ is expected to occur after $1/\rho_{\tau} = \Delta \tau \approx 45$ years. This aligns with the last point of the time series, corresponding to $\tau = 5$ years before 2020. At that time, the rate will be:
    \[
        \rho_{\tau} = \rho_0 + \beta \rho_0^2 \Delta \tau \approx 0.0336 \ [y^{-1}].
    \]
    Thus, the next Canonical Milestone $M_t = 28$ is expected approximately $1 / \rho_{\tau} \approx 30$ years after $\tau = 5$, corresponding to $t = 2045$ CE. Iterating the procedure, the following event $M_t = 29$ is predicted to occur around $t = 2065$ CE. This last prediction is purely speculative. As discussed earlier, predictions far from the last observed point would require updating the parameters using new data. Nevertheless, the prediction for 2065 bypasses the singularity at 2057, illustrating that in the discrete model, the concept of a singularity does not apply. The discrete formulation thus enables forecasting of future events beyond the singularity, addressing the limitations of continuous models.

    As a final remark, note that the scaling factor $\overline{\nu}$ in Eq.~\eqref{eq:model_bio} was not explicitly considered, as we worked directly with the macrodevelopment rate, which does not depend on this factor. In fact, any rescaling of time affects the rate in the same way, ensuring that the results remain unchanged by the choice of $\overline{\nu}$. Nevertheless, its value can still be determined from the data using the relation $\rho_t = \overline{\nu} e^{\beta M_t}$, which directly leads to $\overline{\nu} = \rho_t e^{-\beta M_t}$. By considering the first point in the time series to minimize discretisation errors, we obtain $\overline{\nu} \approx 10^{-10} \ [y^{-1}]$ for CM and $\overline{\nu} \approx 2.2 \cdot 10^{-10} \ [y{-1}]$ for BT.  When modelling the number of milestones, the time-scaling factor $\overline{\nu}$ becomes necessary, since the right-hand side of Eq.~\eqref{eq:model_bio} is always greater than 1, while the actual time scale requires a rate smaller than 1.

    In the case of the world population, we follow the same procedure. Starting from the equation:
    \[
        N_t = \dfrac{1}{N_0^{-1} - \nu_{\gamma} (t - t_0)},
    \]
    we shift the time variable to obtain:
    \[
        N_{\tau} = \dfrac{1}{N_0^{-1} + \nu_{\gamma} (\tau - \tau_0)},
    \]
    where $\tau = 2020 - t$ and $\tau_0 = 2020 - t_0$.
    
    We fit $1/N_\tau$ against time, imposing $\tau_0 = 0$ ($t_0 = 2020$) and $N_0 \approx 7.79 \cdot 10^9$, we obtain:
    \[
        \nu_{\gamma} = (2.01 \pm 0.06) \times 10^{-12} \ [y^{-1}],
    \]
    \[
        \tau^*_{WP} = - \dfrac{1}{N_0 \nu_{\gamma}} = -63.8 \pm 0.6 \ [y], \ (t^*_{WP} \approx 2084 \ [y]).
    \]
    Note the relatively small error in the singularity estimate, which is due to the shorter time interval considered for this dataset compared to the previous ones, as well as the larger number of points, reducing uncertainty. Once again, our predictions differ from previous literature due to the use of different data and time periods. 

    Due to the smaller time scale considered, in this case we can safely put $\overline{\nu} = 1 \ [y^{-1}]$~\cite{cortes2022biocosmology}. As a consequence, the macrodevelopment parameter $\gamma$ is equal to $\nu_{\gamma}$, without dimension.
    This parameter can be interpreted as a birth rate, indicating the fraction of potential interactions between individuals that lead to successful reproduction. Considering the discrete version of the model:
    \[
        N_{t+1} - N_t \approx \gamma N_t^2,
    \]
    we find, for example, that with the current population size of approximately $8 \cdot 10^9$, the predicted increase in the next year is about $\gamma N_t^2 \approx 1.2 \cdot 10^8$. This estimate aligns with the actual population growth observed between 2020 and 2023, which averaged around $0.75 \cdot 10^8$ per year.
    
    For GDP growth, we use the equation:
    \[
        K_{\tau} \approx \dfrac{1}{\left( K_0^{-1/2} + \nu_{\delta}(\tau-\tau_0) \right)^2}.
    \]
    By fitting the linear relation between $1 / K_\tau^2$ and time, and using the initial conditions $\tau_0 = 0$ ($t_0 = 2018$) and $K_0 = 3.402405 \cdot 10^6$, we obtain:
    \[
        \nu_{\delta} = (1.96 \pm 0.03) \times 10^{-5} \ [y^{-1}],
    \]
    \[
        \tau^*_{GDP} = - \dfrac{1}{\sqrt{K_0} \nu_{\delta}} = -27.7 \pm 0.4 \ [y], \ (t_{GDP}^* \approx 2048 \ [y]).
    \]
    As for the world population, in this case the time scaling factor can be put equal to $\overline{\nu} = 1 \ [y^{-1}]$.
    The dimensionless macrodevelopment parameter $\delta = (1.96 \pm 0.03) \times 10^{-5}$ represents the rate of GDP growth. Using the discrete dynamics:
    \[
        K_{t+1} - K_t \approx \delta K_t^{3/2},
    \]
    we estimate that in the year following 2018, GDP should grow by approximately $\delta K_t^{3/2} \approx 1.2 \cdot 10^5$ billions of USD dollars. The actual GDP growth observed over the years (2019-2022) was about 3\% per year. Given that $K_t \approx 3.40 \times 10^6$ billions of USD dollars at $t_0 = 2018$, this corresponds to an annual increase of approximately $1.0 \times 10^5$ billions of USD dollars, aligning well with our estimate. 

    Finally, the number of patents as a function of the number of technological codes is given by:
    \[
        P_C = \dfrac{e^{\beta C}}{\beta} - \dfrac{e^{\beta C_0}}{\beta} + P_0.
    \]
    The equation for patents (Eq.~\eqref{eq:patent}) is dimensionless, so in this case, there is no need to account for a dimensional factor.
    From the exponential fit, and setting $C_0$ and $P_0$, we find $\beta = (4.18 \pm 0.02) \times 10^{-5}$. In this case, there is no singularity, even in the continuous version of the model.
    
    The macrodevelopment parameter $\alpha = e^{\beta} - 1$ reflects the growth rate of the number of patents relative to the number of available technological codes. The discrete dynamics can be expressed as:
    \[
        P_{C+1} - P_C = \sum_{i=1}^{C} \alpha^i \binom{C}{i}.
    \]
    By substituting the actual number of technological codes $C = 160586$ in the year 2020, we can compute the expected number of patents introduced when $C$ increases by one, i.e., when a new technological code is created. This results in approximately $P_{C+1} - P_C \approx 821$ new patents. In other words, at this stage of the system's evolution, it takes roughly $821$ patents to introduce a single new technological code.

\end{document}